\documentclass[a4paper]{jpconf}
\usepackage{graphicx}
\usepackage{amsmath}
\usepackage{amsfonts}
\usepackage{amssymb}
\begin{document}
\title{The proton radius and its relatives - \\much ado about nothing?}

\author{Ulf-G. Mei{\ss}ner}

\address{HISKP and BCTP, Bonn University, D-53115 Bonn, Germany\\
  IKP-3 and IAS-4, Forschungszentrum J\"ulich, D-52425 J\"ulich, Germany\\
  Tbilisi State University, 0186 Tbilisi, Georgia}

\ead{meissner@hiskp.uni-bonn.de}

\begin{abstract}
I summarize the dispersion-theoretical analysis of the nucleon electromagnetic form factors.
Special emphasis is given on the extraction of the proton charge radius and its relatives,
the proton magnetic radius as well as the neutron magnetic radius. Some recent work on the
hyperfine splitting in leptonic hydrogen and on radiative corrections to muon-proton
scattering is also discussed. Some views on future studies are given.
\end{abstract}

\section{Introduction}

The measurement of the 2S-2P energy splitting in muonic hydrogen by the CREMA collaboration
with $10^{-5}$ relative accuracy~\cite{Pohl:2010zza,Antognini:2013txn} started the
so-called ``proton radius puzzle'', as the then accepted value from electron-proton scattering
was significantly larger, see e.g. \cite{A1:2013fsc}, contrasting the ``small radius'' $r_E^p \simeq
0.84\,$fm with the ``large'' one,  $r_E^p \simeq 0.88\,$fm. This stirred lots of activities, both on the
experimental as well as on the theoretical side, see e.g. the recent reviews~\cite{Karr:2020wgh,Gao:2021sml}.
However, as I will show in this talk, the dispersion-theoretical analysis of the $ep$ scattering
data and the available data for the form factors from data in the timelike region always led
to the small radius and continues to do so. In fact, the radius puzzle was anticipated in 2007 in
Ref.~\cite{Belushkin:2006qa}, where it was shown that the DR analysis of $ep$ scattering
could not accommodate the then accepted radius from the Lamb shift in electronic hydrogen of
$r_E^p =0.88-0.89\,$fm.
Without going into further details of this intriguing story, that is, without telling about all
the twists and turns (or: confusions and errors) that appeared over time, I will  illuminate some
developments that were not appreciated by large parts of the community for a long time. More precisely,  
what I will show here is that the
powerful approach of dispersion relations (DRs), that allows one to include all possible physics constraints,
is arguably the best method to analyze the data that are sensitive to the nucleons electromagnetic
form factors and thus the various sizes of the nucleon as given by the electric and magnetic radii of the
proton and the neutron.

\section{Formalism}
\label{sec:form}

First, let me define  the nucleon electromagnetic form factors. For the
dispersive analysis it is mandatory to consider protons and neutrons together,
for details see the review~\cite{Lin:2021umz}. These form factors are given
by the matrix element of the electromagnetic current $j_\mu^{\rm em}$ sandwiched between
nucleon states,
\begin{equation}
\langle p' | j_\mu^{\rm em} | p \rangle = \bar{u}(p')
\left[ F_1 (t) \gamma_\mu +i\frac{F_2 (t)}{2 m} \sigma_{\mu\nu} q^\nu \right] u(p)\,,
\end{equation}
with $m$ the nucleon mass (either proton or neutron), $u(p)$ a conventional nucleon
spinor and $t=(p'-p)^2$  the four-momentum
transfer squared. In the space-like region, one often
uses the variable $Q^2=-t>0$.
The form factors are normalized as
$F_1^p(0) = 1\,$,  $F_1^n(0) = 0\,$, $F_2^p(0) =  \kappa_p\,$ and $F_2^n(0) = \kappa_n\,$,
with $\kappa_p=1.793$ and $\kappa_n=-1.913$ the anomalous magnetic moment of
the proton and the neutron, respectively. Also used are the Sachs (electric and magnetic) form factors,
given by $G_{E}(t) = F_1(t) - \tau F_2(t)$, $G_{M}(t) = F_1(t) + F_2(t)$, 
with $\tau = -t/(4 m^2)$. The proton charge radius and the corresponding magnetic
radius follow as
\begin{equation}
\label{eq:rp}
 (r_E^p)^2 = 6 \frac{dG_p^E(t)}{dt}\Big|_{t=0}~, \quad  (r_M^p)^2 = \frac{6}{\mu_p} \frac{dG_p^M(t)}{dt}\Big|_{t=0}~,
\end{equation}
with $\mu_p=1+\kappa_p$. The neutron radii are defined similarly, note, however, that $G_E^n(0) = 0$.

\begin{figure}[t]
\centering\includegraphics*[width=0.75\linewidth,angle=0]{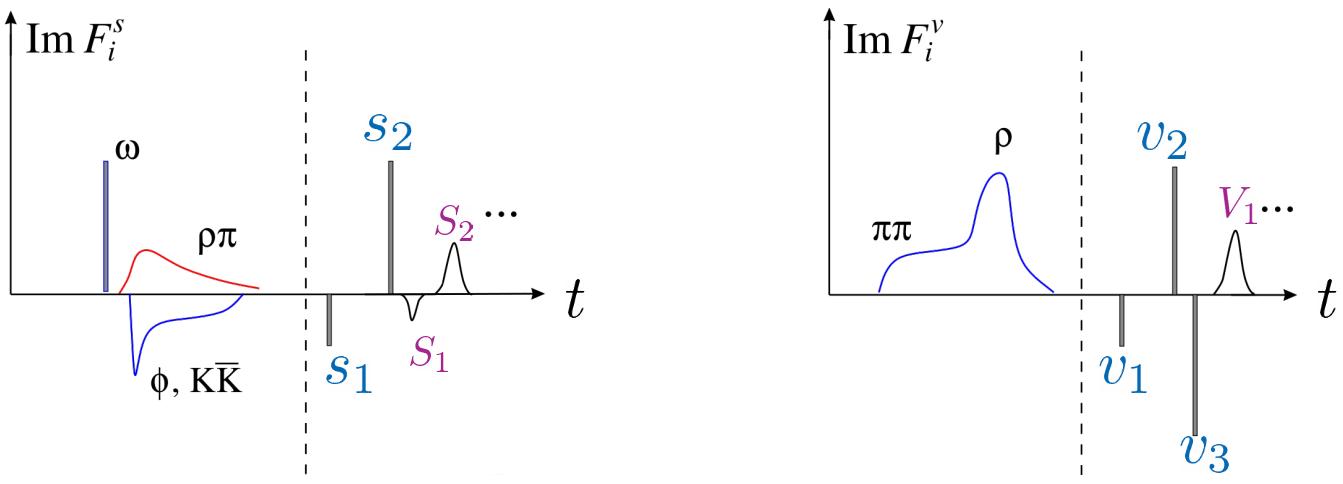}~~~~~
\caption{\label{fig:spec}Sketch of the isoscalar (left) and isovector (right) spectral function.
In the left panel, the $\omega$ and $\phi$ mesons are relevant together with the  $\pi\rho$ and $K\bar{K}$ continua,
while $s_1, s_2, ...$ are narrow and $S_1, S_2, ...$ are broad effective poles. In the right panel,
the $\pi\pi$ continuum not only generates the $\rho$ but is also visibly enhanced on the left shoulder of the $\rho$.
Further,  $v_1, v_2, v_3, ...$ are narrow and $V_1, ...$ are  broad effective poles.}
\vspace{-3mm}
\end{figure}

Next, we turn to the dispersive analysis of the nucleon electromagnetic form factors.
For a generic form factor $F(t)$, one writes down an unsubtracted dispersion
relation of the form:
\begin{equation}
F(t) = \frac{1}{\pi} \, \int_{t_0}^\infty \frac{{\rm Im}\, 
F(t')}{t'-t-i\epsilon}\, dt'\, .
\label{eq:disp} 
\end{equation}
Here, $t_0$ is the threshold of the lowest cut of $F(t)$ and the $i\epsilon$ defines the
integral for values of $t$ on the cut. In fact, in the isospin basis, $t_0 = 4M_\pi^2$ in the
isovector and $t_0 = 9M_\pi^2$ in the isoscalar channel, respectively. The imaginary part ${\rm Im}\, F$,
the so-called {\em spectral function}, encodes the constraints from analyticity and unitarity besides
other important physics.
These spectral functions are given in terms of continua, narrow vector
meson poles as well as broad vector mesons. In the isovector case, the spectral function can
be reconstructed up to about $t \simeq 1\,$GeV$^2$ from data on pion-nucleon scattering and
the pion vector form factor, as pioneered in Ref.~\cite{Frazer:1959gy}
and most precisely done in Ref.~\cite{Hoferichter:2016duk}. This
in fact not only generates the $\rho$-meson but also an important enhancement on the left
shoulder of the $\rho$, that is of utmost importance to properly describe the nucleon isovector
radii~\cite{Hohler:1974eq}.
In the isoscalar spectral function, the $\omega$-meson represents the lowest contribution. It is not
affected by uncorrelated three-pion exchange~\cite{Bernard:1996cc}.
Further up, in the region of the $\phi$-meson, there
is a strong competition between $K\bar{K}$ and $\pi\rho$ effects, which to some extent suppresses
this part of the spectral function. For momenta above $t \simeq 1\,$GeV$^2$, effective narrow poles
represent the physics at higher energies. To describe the observed oscillations
of the cross sections for $e^+e^-\to p\bar{p}$ and $e^+e^-\to n\bar{n}$  in the timelike region, 
additional broad poles are required. The spectral functions are further constrained by the
normalizations of the form factors as well as the perturbative QCD behavior,
$F_1 (t) \sim 1/t^2$ and $F_2(t) \sim 1/t^3$.  A cartoon of the spectral functions is
given in Fig.~\ref{fig:spec}.

The spectral functions are determined from a fit to the world data set on electron-proton scattering
as well as the reactions $e^+e^- \leftrightarrow \bar{p}p, \bar{n}n$, the latter giving the form factors
in the timelike region. The fit parameters are the vector meson masses (except for the $\omega$
and the $\phi$) and the residua as well as the widths for the broad poles. There are two sources
of uncertainties that need to be accounted for. First, the statistical error is obtained using
a bootstrap procedure and second, the systematic error is calculated from varying the number
of vector meson poles so that the total $\chi^2$ does not change by more than 1\%~\cite{Hohler:1976ax}.
A detailed description of these methods is given in the review~\cite{Lin:2021umz}.

\section{Results for the form factors}

\begin{figure}[h!] 
\includegraphics*[width=0.45\linewidth,angle=0]{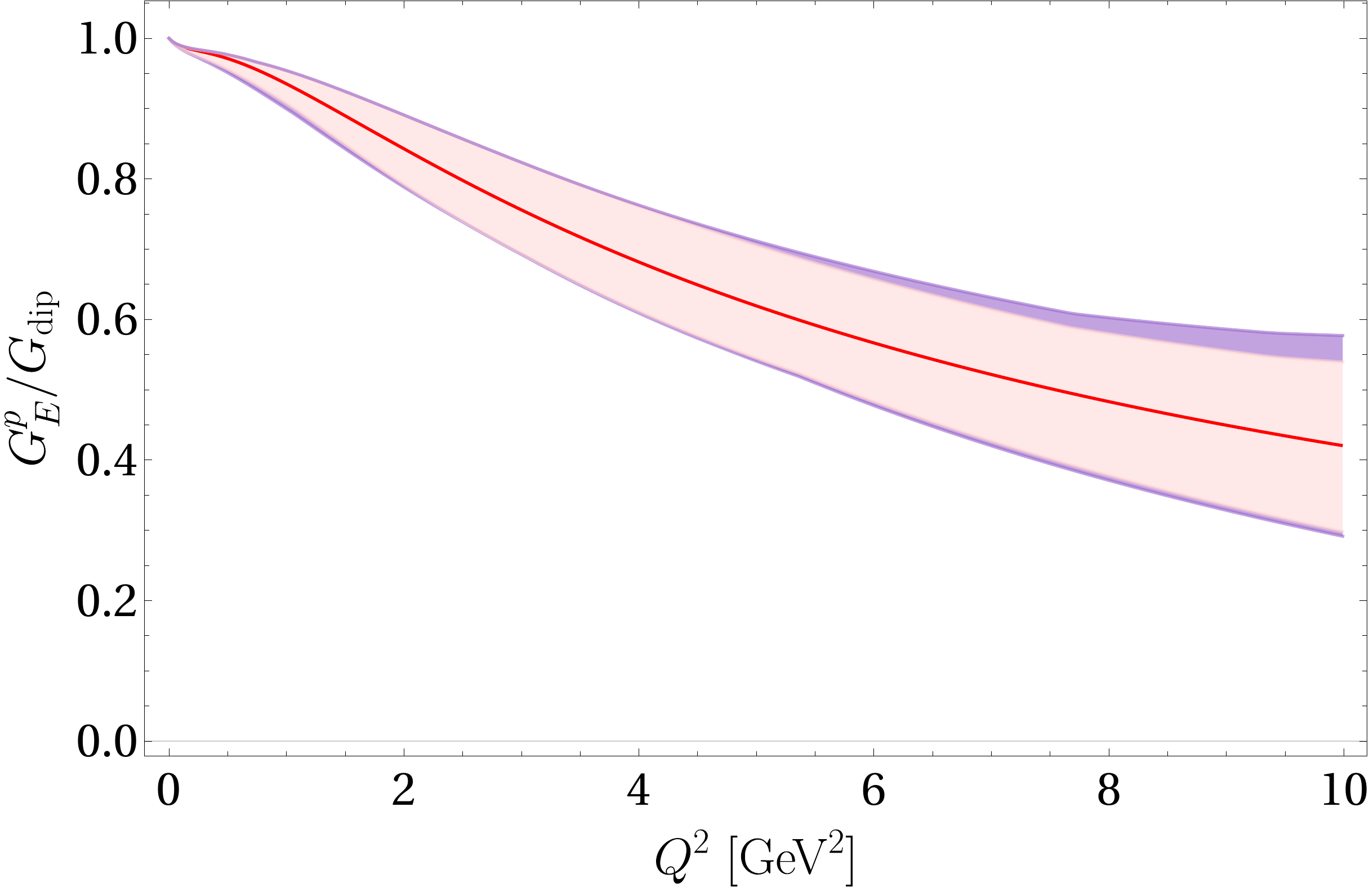}~~~~
\includegraphics*[width=0.45\linewidth,angle=0]{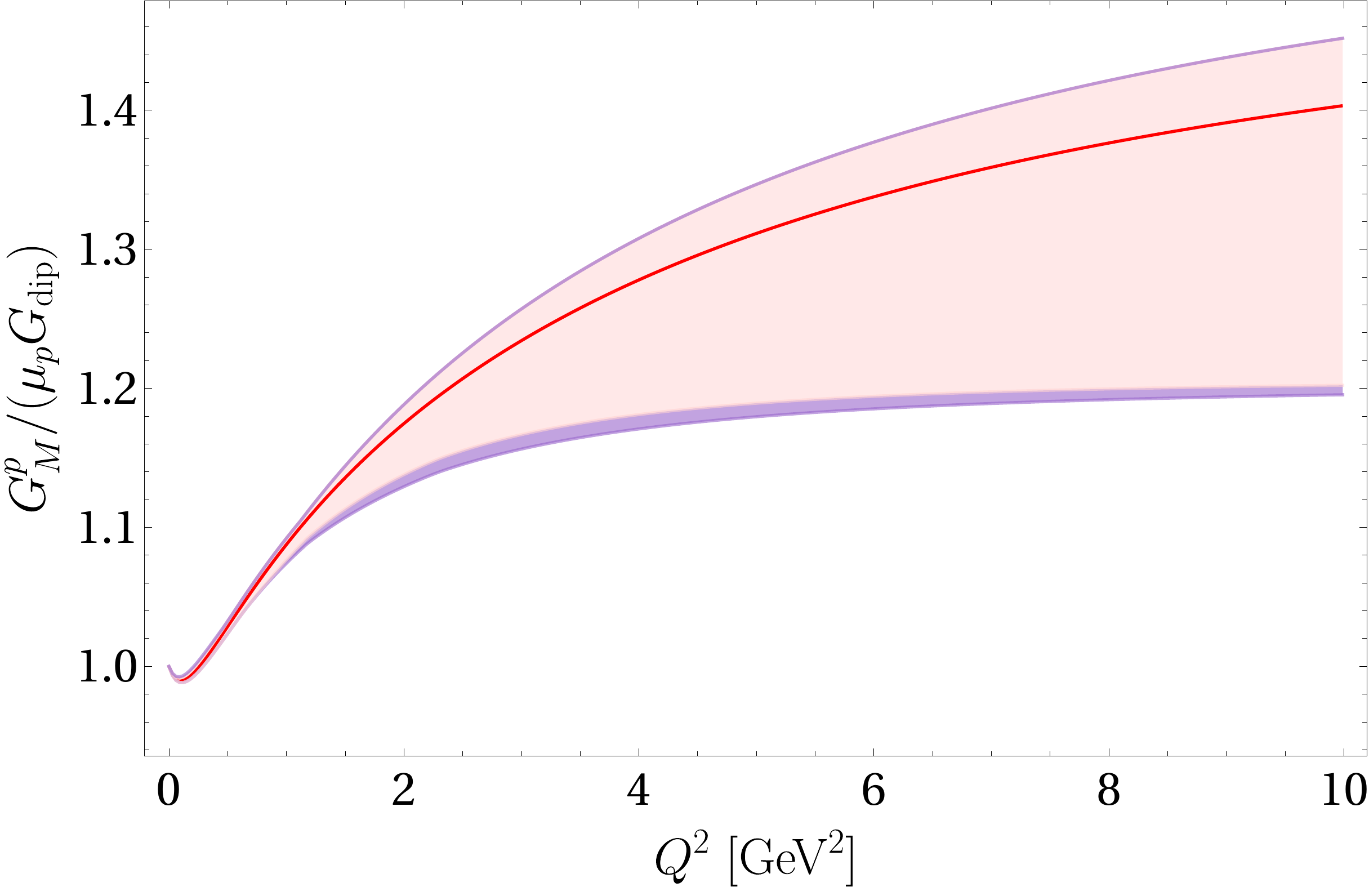} 
\caption{Electric (left panel) and magnetic (right panel) form factor of the proton
  from Ref.~\cite{Lin:2021xrc}
  divided by the canonical dipole form factor are shown by the red lines. The light red band is the
  statistical uncertainty and the purple band shows the systematic error added in quadrature. }
\label{fig:Gp}
\vspace{-3mm}
\end{figure}

Fitting to the world data basis of data in the spacelike and the timelike regions (about 1800 data points),
the form factors are obtained with good precision. Details of these fits are given in Ref.~\cite{Lin:2021xrc}.
The electric and magnetic form factors of the proton in the spacelike region from Ref.~\cite{Lin:2021xrc}
normalized to the canonical dipole form, $G_{\rm dip}(Q^2) = (1+Q^2/0.71\,{\rm GeV}^2)^{-2}$, are shown in
Fig.~\ref{fig:Gp} together with their statistical and systematic uncertainties. It is interesting to note that
the form factor ratio $\mu_p G_E^p(Q^2)/G_M^p(Q^2)$ measured at Jefferson Lab seems to level off with
increasing momentum transfer, somewhat disfavoring a zero crossing. To settle this issues certainly requires
more measurements extending to $Q^2 \simeq 10\,$GeV$^2$.
Predictions for the upcoming PRad-II experiment at Jefferson Lab are made in Ref.~\cite{Lin:2021cnk}.

The form factors in the timelike region are complex-valued quantities. In most experiments,
the modulus of the effective form factor $|G_{\rm eff}(t)|$ is extracted, which is a linear
combination of $|G_E(t)|$ and $|G_M(t)|$, see e.g.~\cite{Lin:2021umz,Pacetti:2014jai} for definitions.
There are clearly visible oscillations, which in the DR approach are generated by the interference
of the broad poles. However, there are other mechanisms proposed to generate such structures,
see e.g. Refs.~\cite{Bianconi:2015owa,Bianconi:2015vva,Lorenz:2014yda,Tomasi-Gustafsson:2020vae,Yang:2022qoy}.

\begin{figure}[h!] 
\includegraphics*[width=0.45\linewidth,angle=0]{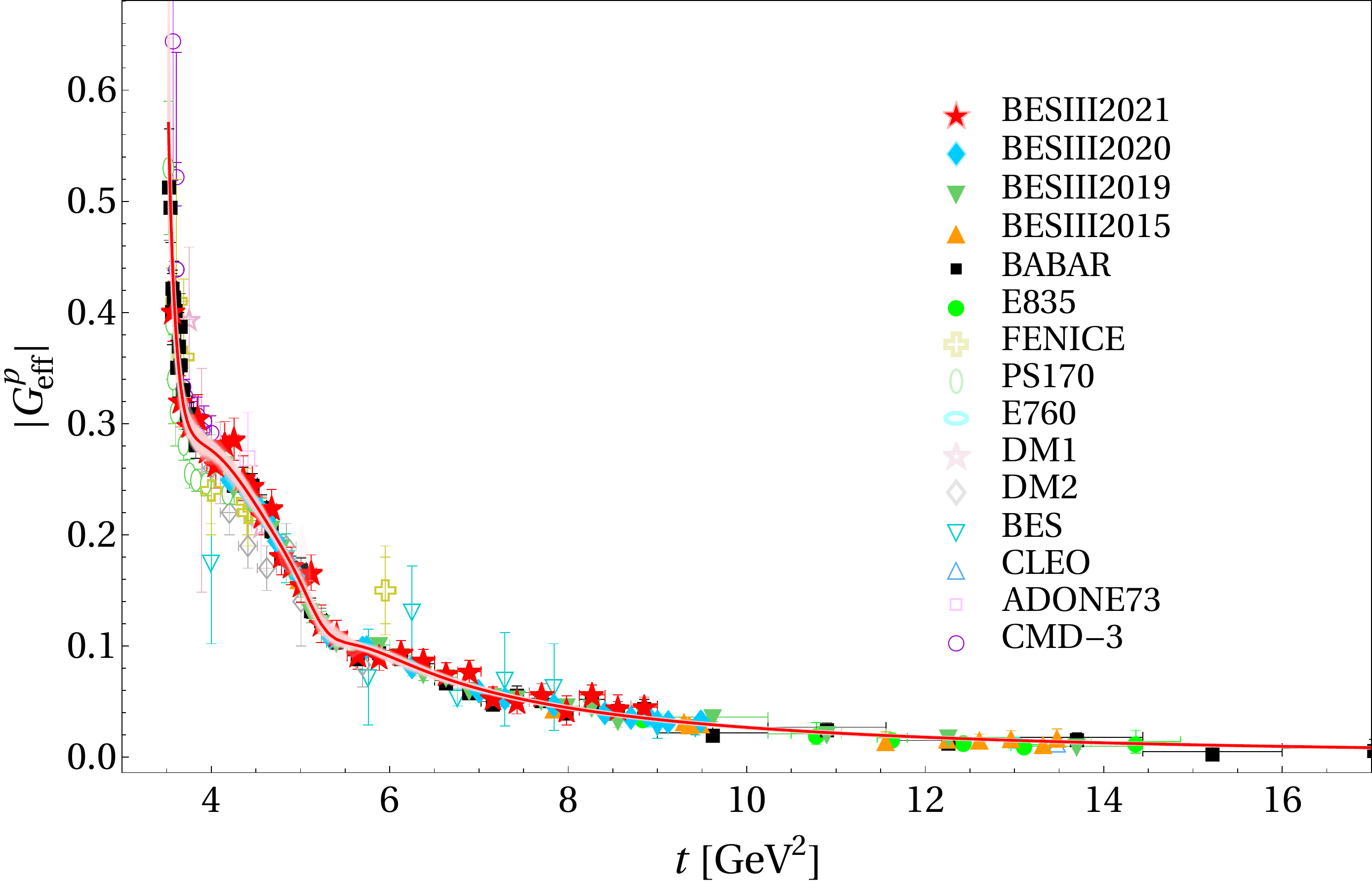}~~~~
\includegraphics*[width=0.45\linewidth,angle=0]{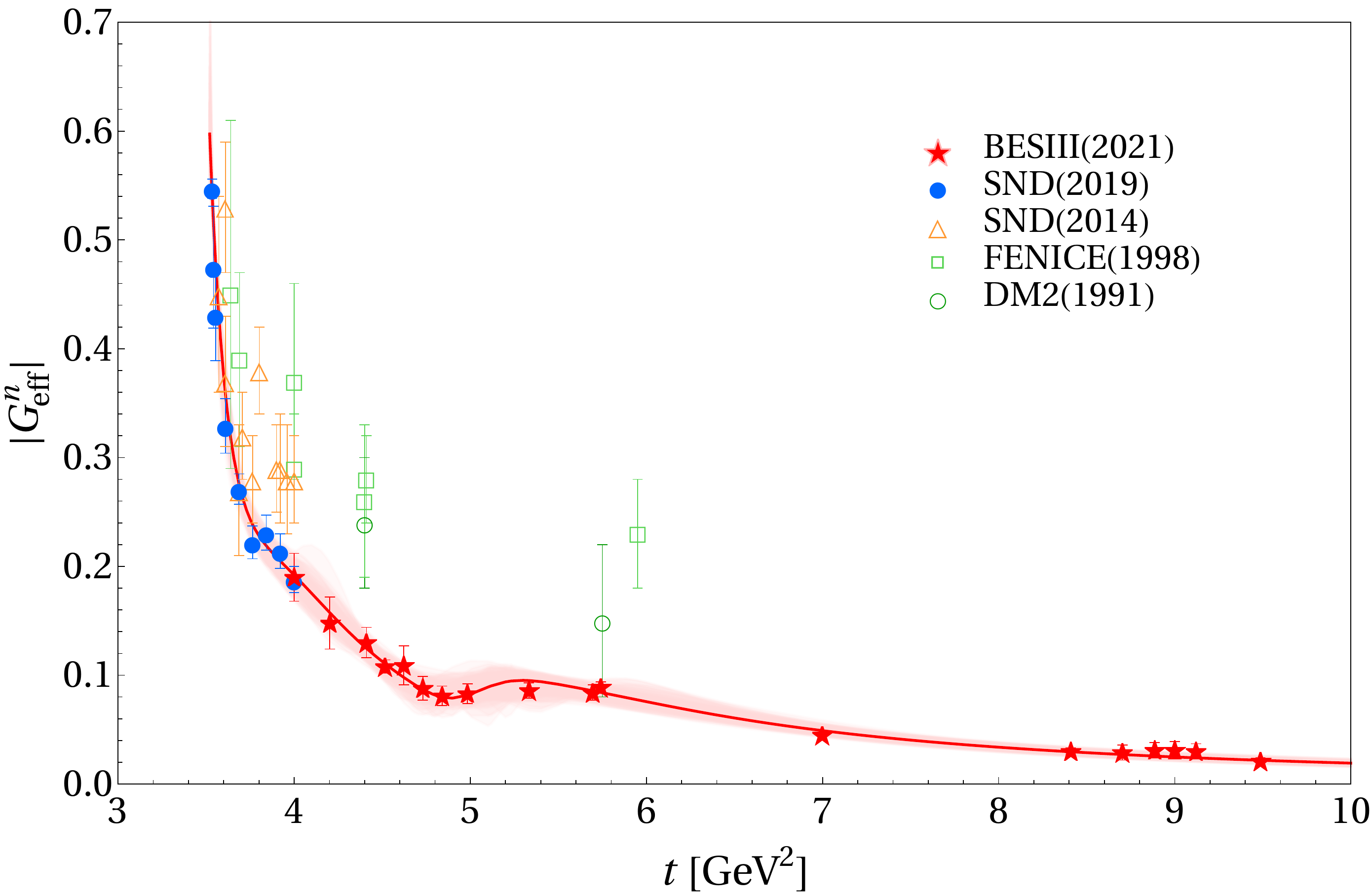} 
\caption{Oscillations in the  proton  (left panel) and neutron (right panel) effective form factor 
  in the timelike region from the complete fit to all data~\cite{Lin:2021xrc}. Closed symbols are
  fitted and open symbols are not fitted. The complete data basis is given in Ref.~\cite{Lin:2021xrc}.}
\label{fig:Gosc}
\vspace{-3mm}
\end{figure}

\section{Results for the radii}

\begin{figure}[t]
\includegraphics[width=32pc]{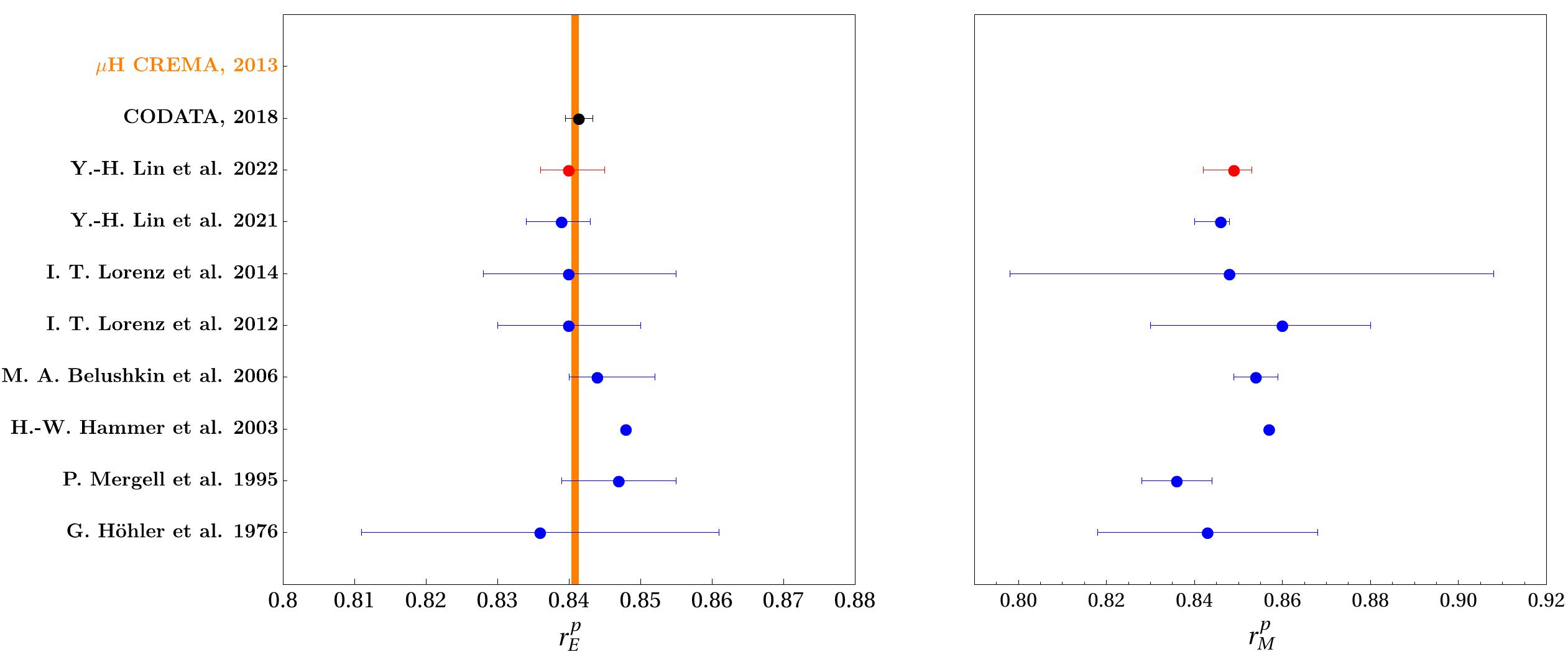}
\caption{\label{fig:comp}DR determinations of the proton charge radius (left panel) and the
  proton magnetic radius (right panel). Also shown are the CREMA result and the CODATA 2018 value for $r_E^p$.
  Values are in [fm]. References from top to bottom:
  \cite{Antognini:2013txn,Tiesinga:2021myr,Lin:2021xrc,Lin:2021umk,Lorenz:2014yda,Lorenz:2012tm,Belushkin:2006qa,Hammer:2003ai,Mergell:1995bf,Hohler:1976ax}.}
\vspace{-3mm}
\end{figure}

From the just discussed form factors, the extracted proton radii are~\cite{Lin:2021xrc} 
\begin{equation}
r_E^p = 0.840^{+0.003}_{-0.002}{}^{+0.002}_{-0.002}~{\rm fm}~, ~~~
r_M^p = 0.849^{+0.003}_{-0.003}{}^{+0.001}_{-0.004}~{\rm fm}~,
\end{equation}
where the first error is statistical and the second one systematic. This value of the proton
charge radius is consistent with recent experimental determination as shown in Tab.~\ref{tab:rEp}
and the present CODATA value of $0.8414(19)\,$fm. Other recent determinations of the proton charge
radius give $r_E^p = 0.842(10)\,$fm ~\cite{Alarcon:2020kcz}, $0.847(8)\,$fm~\cite{Cui:2021vgm} and
$0.852(9)\,$fm~\cite{Atac:2021} and for the magnetic radius  $r_M^p = 0.850(10)\,$fm ~\cite{Alarcon:2020kcz},
$0.739(41)\,$fm~\cite{Borah:2020gte} and $0.817(27)\,$fm~\cite{Cui:2021skn}. For more discussion on
these determinations, see Ref.~\cite{Lin:2021umk}. As further shown in Fig.~\ref{fig:comp}, the
proton charge and magnetic radius have been rather stable since the 1976 Karlsruhe DR \cite{Hohler:1976ax},
with $r_M^p > r_E^p$ always. 

\begin{table}
  \caption{Recent determination of the proton charge radius from the electronic Lamb shift
  in hydrogen (H) and $ep$ scattering compared to the most recent DR result.\label{tab:rEp}}
\begin{center}
\begin{tabular}{|c|c|c|c|}
\hline
$r^p_E$ [fm] & year &  method & Ref. \\
\hline
0.877(13)  & 2018 & H Lamb shift &\cite{Fleurbaey:2018fih}\\
0.833(10)  & 2019 & H Lamb shift &\cite{Bezginov:2019mdi}\\
0.8482(38) & 2020 & H Lamb shift &\cite{Grinin:2020}\\
0.8584(51) & 2021 & H Lamb shift &\cite{Brandt:2021yor}\\
\hline
0.831(7)(12) & 2019& $ep$ scattering & \cite{Xiong:2019umf}\\
\hline
0.840(3)(2) & 2022& DR & \cite{Lin:2021xrc}\\
\hline
\end{tabular}
\end{center}
\vspace{-3mm}
\end{table}  

\begin{figure}[h]
\includegraphics[width=18pc]{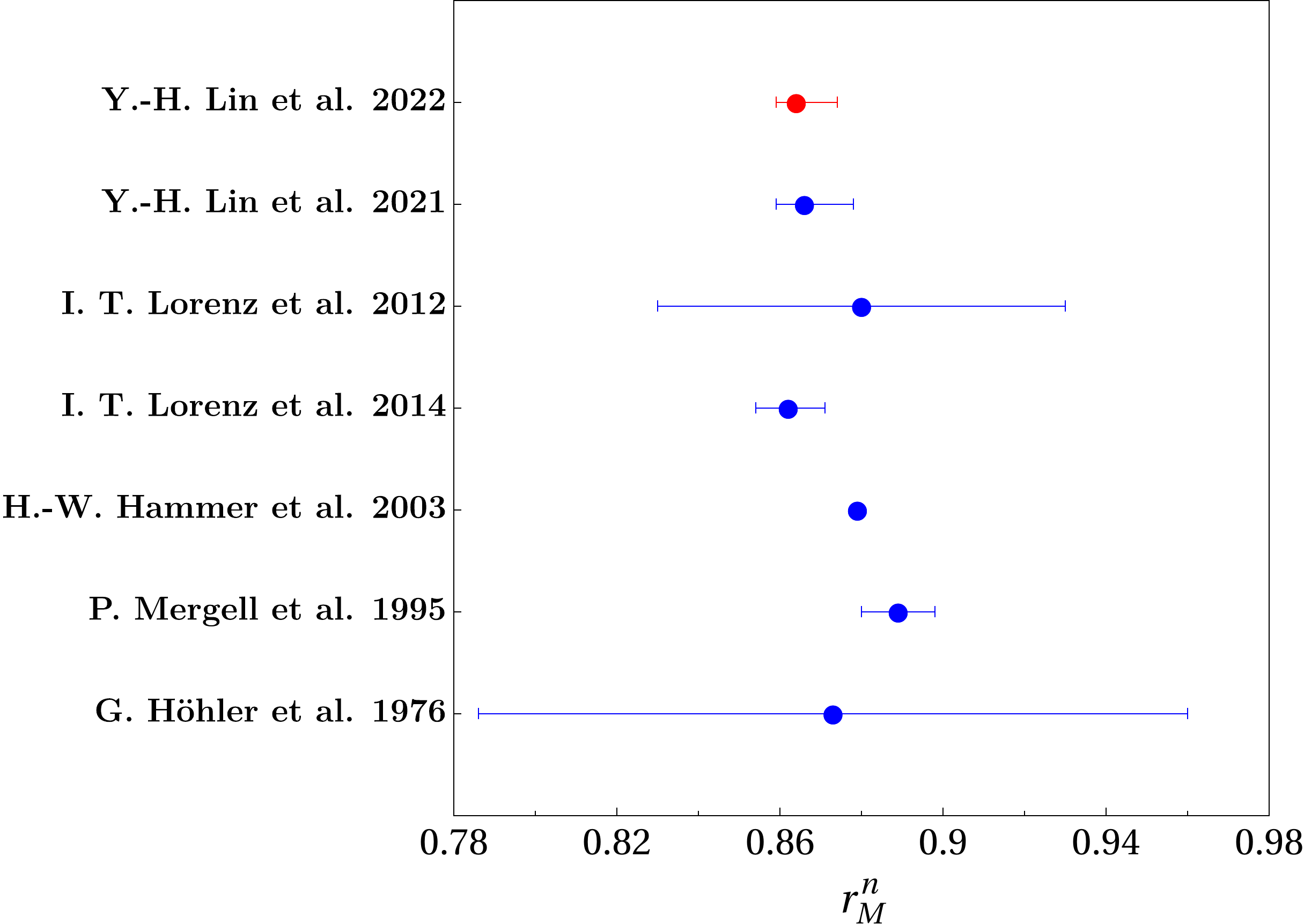}\hspace{2pc}%
\begin{minipage}[b]{16pc}\caption{\label{fig:rnm}Extraction of the neutron magnetic radius
    $r_M^n$ in units of fm  based on dispersion-theoretical analyses since 1976. References
  as in Fig.~\ref{fig:comp}.}
\end{minipage}
\end{figure}

Let me now consider the neutron radii. The charge radius squared is used as input in most DR analyses,
with the exception of Ref.~\cite{Belushkin:2006qa}, where it was also extracted (and came out consistent
with other determinations). In the most
recent analysis, the precise value extracted from $ed$ scattering using chiral nuclear EFT was used,
$(r_E^n)^2 = -0.105^{+0.005}_{-0.006}\,$fm$^2$~\cite{Filin:2020tcs}. The extracted magnetic radius for the
neutron is
\begin{equation}
r_M^n = 0.864^{+0.004}_{-0.004}{}^{+0.006}_{-0.001}~{\rm fm}~.
\end{equation}
The uncertainties are a bit larger than for the proton, but in any case, the neutron magnetic
radius is the largest of all, $r_M^n >r_M^p > r_E^p$, and it has been equally constant over
time as shown in Fig.~\ref{fig:rnm}. For comparison, Ref.~\cite{Borah:2020gte} gives
$r_M^n = 0.878(79)$~fm. I mention that the bump-dip structure seen at low $Q^2$ in $G_M^n(Q^2)$ in
some form factor extractions is inconsistent with the strictures from analyticity and unitarity, as such
a structure would require a vector meson with a mass of about 400~MeV that is excluded by the spectral
functions shown above.

Finally, it is interesting to compare the DR results with recent lattice QCD determinations.
In Tab.~\ref{tab:lat}, the isovector radii are displayed as these are free of disconnected diagrams.
Only simulations at the physical pion mass are considered. While the isovector electric
radius appears to converge to the DR result, the situation is less clear in the isovector magnetic case.
Clearly, more precise lattice studies need to be performed.

\begin{table}[h]
  \caption{Recent determination of the nucleon isovector radii from lattice QCD simulations with
    physical pion masses. \label{tab:lat}} 
\begin{center}
\begin{tabular}{|l|c|c|}
\hline
                   &  $r_E^V$ [fm]   &  $r_M^V$ [fm] \\
\hline
DR ~\cite{Lin:2021xrc}         &  0.900(2)(2)    &   0.854(1)(3)  \\
\hline
Lattice/Mainz (2021) \cite{Djukanovic:2021cgp}  &  0.894(14)(12)  &   0.813(18)(7) \\
Lattice/ETMC  (2020) \cite{Alexandrou:2020aja} &  0.827(47)(5)    &    ----       \\
Lattice/PACS  (2019) \cite{Shintani:2018ozy} &  0.785(17)(21)  &   0.758(33)(286) \\
Lattice/MIT   (2018) \cite{Hasan:2017wwt} &  0.787(87)      &    ----        \\
\hline
\end{tabular}
\end{center}
\vspace{-5mm}
\end{table}

\section{Remarks on the hyperfine-spitting in leptonic hydrogen}

Laser spectroscopy of muonic hydrogen ($\mu$H), an atom formed by a negatively charged
muon and a proton, represents an excellent pathway to investigate low-energy properties of the proton.
While the 2S-2P energy splitting is sensitive to electric properties of the proton as the
proton charge radius as discussed before, the hyperfine splitting (HFS) is sensitive also to magnetic properties of
the proton as it arises from the interaction between the proton and muon magnetic moments.
For recent reviews, see Refs.~\cite{Peset:2021iul,Antognini:2022xoo}.
For the HFS, the leading proton structure contribution is given by the two-photon-exchange
contribution $\Delta E^{2\gamma} $, which is conventionally divided into a Zemach radius
contribution $\Delta_Z^{\mu{\rm H}}$, a recoil contribution $\Delta_\mathrm{recoil}^{\mu{\rm H}}$ and a
polarizability contribution $\Delta_\mathrm{pol}^{\mu{\rm H}}$
\cite{Carlson:2008ke,Carlson:2011af,Tomalak:2017lxo,Tomalak:2017owk,Faustov:2006ve,Hagelstein:2015egb}.
The Zemach contribution $\Delta_Z^{\mu{\rm H}}$ that accounts for the elastic part of the two-photon exchange
contribution can be expressed through the electric  and  magnetic  Sachs form factors
of the proton~\cite{Zemach:1956zz}
\begin{equation}
\label{eq:Zemach}
\Delta_Z  =-2Z\alpha m_r  \, r_\mathrm{Z}^p~, \quad
r_\mathrm{Z}^p = - \frac{4}{\pi} \int_0^\infty \frac{dQ}{Q^2}
  \left[\frac{G_E^p(Q^2) G_M^p(Q^2) }{1+\kappa_p} -1\right]~,
\end{equation}
where $Z$ is the atomic number, $\alpha = e^2/(4\pi)$ the electromagnetic fine-structure constant,
$\mu_r$ the reduced mass of the lepton-proton system and $r_\mathrm{Z}^p$ is the Zemach radius.
The so-called recoil contribution, which more precisely is the recoil correction to
the Zemach contribution, can also be described solely by form factors. In addition to
the Sachs form factors in this case also the Dirac and Pauli form factors enter~\cite{Antognini:2022xoo}:
\begin{eqnarray}
  \Delta_\mathrm{recoil} &=&\frac{Z \alpha}{\pi (1+\kappa_p)}\int_0^\infty \frac{dQ}{Q}
  \Bigg\{\frac{G_M^p(Q^2)}{Q^2}\frac{8m_l m_p}{v_l+ v_p} \left(2F_1^p(Q^2)+\frac{F_1^p(Q^2)
    +3F_2^p(Q^2)}{(v_l+1)(v_p+1)}
  \right)\nonumber\\ &&-\frac{8m_r G_M^p(Q^2)G_E^p(Q^2)}{Q}-\frac{m_l (F_2^p)^2(Q^2)}{m_p}
  \frac{5+4v_l}{(1+v_l)^2}\Bigg\}~,
  \label{Deltarecoilintegrand}
\end{eqnarray}
%
where $v_p=(1+4m_p^2/Q^2)^{1/2}$, $v_l= (1+4m_l^2/Q^2)^{1/2}$ and $m_p \,(m_l)$ is the proton (lepton) mass.
A precise evaluation of $\Delta_\mathrm{recoil}^{\mu{\rm H}}$  
is timely given the ongoing experimental efforts carried out by three collaborations
that aim at the HFS in $\mu$H~\cite{Amaro:2021goz, Kanda:2020mmc, Pizzolotto:2020fue} 
with relative accuracies ranging from 1 to 10~ppm.

\begin{figure}[h]
\includegraphics[width=20pc]{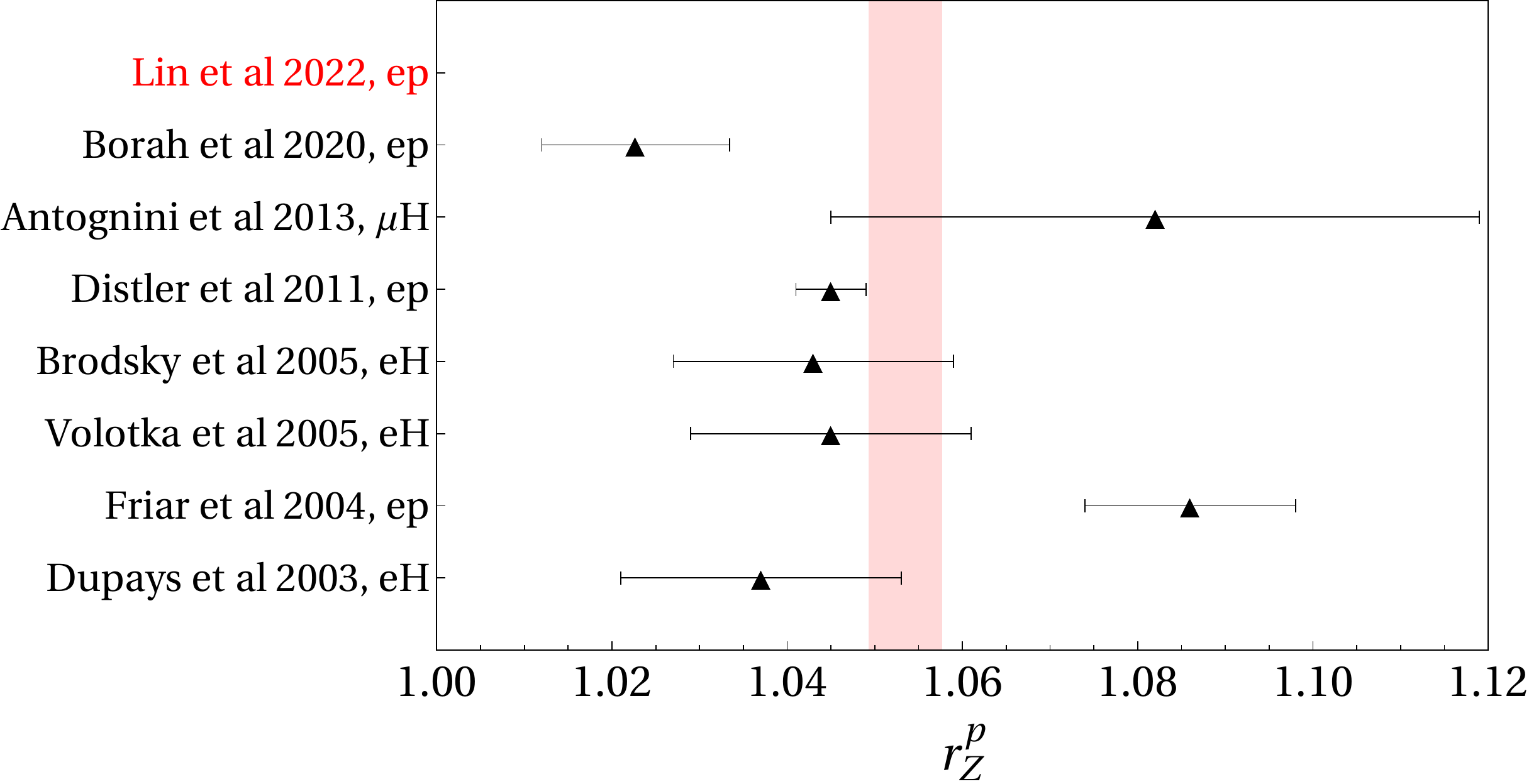}\hspace{2pc}%
\begin{minipage}[b]{16pc}\caption{\label{fig:Zr}Extraction of the Zemach radius (in units of fm)
    from various methods ($ep$ scattering, electronic hydrogen and muonic hydrogen).
    References from top to bottom:
    \cite{Lin:2021xrc,Borah:2020gte,Antognini:2013txn,Distler:2010zq,Brodsky:2004ck,Volotka:2004zu,Friar:2003zg,Dupays:2003zz}.\label{fig:rZ}}
\end{minipage}
\end{figure}

Based on the form factors from~\cite{Lin:2021xrc}, the Zemach radius is:
\begin{equation}
r_Z^p = 1.054^{+0.003}_{-0.002}{}^{+0.000}_{-0.001}\,{\rm fm}~,
\end{equation}  
which is compared to earlier determinations in Fig.~\ref{fig:rZ}.

Consider now the calculation of the recoil correction defined in Eq.~(\ref{Deltarecoilintegrand})
using the  form factors from~\cite{Lin:2021xrc}. For the $\mu$H system, this gives~\cite{Antognini:2022xqf}
\begin{equation}
  \Delta_\mathrm{recoil}^{\mu{\rm H}} =  (837.6^{+1.7}_{-1.0}{}^{+2.2}_{-0.1}) \times 10^{-6}
  = (837.6^{+2.8}_{-1.0})  \times 10^{-6} = (837.6^{+2.8}_{-1.0})~{\rm ppm}~,
\end{equation}  
with the first/second error is the statistical/systematic uncertainty. These
errors are a few per~mille, so that this
can be considered as a high-precision determination. Compared with the most recent value
from Ref.~\cite{Tomalak:2017lxo}, $\Delta_\mathrm{recoil}^{\mu{\rm H}} = 844(7) \times 10^{-6}$,
these numbers agree  within errors but the new result is more precise.
The analogous value for regular hydrogen ($e$H) is~\cite{Antognini:2022xqf}
\begin{equation}
  \Delta_\mathrm{recoil}^{e\rm H} =   (526.9^{+1.1}_{-0.3}{}^{+1.3}_{-0.2}) \times 10^{-8}
  = (526.9^{+1.7}_{-0.4}) \times 10^{-8}~,
\end{equation}    
which is, as expected, two orders of magnitude smaller but with comparable
uncertainties as in the $\mu$H case.  Again, the corresponding value from
Ref.~\cite{Tomalak:2017lxo}, $\Delta_\mathrm{recoil}^{e\rm H} = 532.8(4.9)\times 10^{-8}$,
is about 1\% larger but is also a bit less precise. For further discussion, see Ref.~\cite{Antognini:2022xqf}.

\section{Remarks on muon-proton scattering}

Elastic muon-proton scattering at low momentum transfers offers an alternative
method to measure the proton charge radius $r_E^p$. Any deviation from the value measured
in $ep$ scattering would challenge the concept
of lepton-flavor universality, which is a cornerstone of the so successful Standard Model of particle
physics, that was put into question in recent experiments on certain decay modes of B mesons.
There are experiments that are pursuing such proton radius measurements, namely MUSE
at PSI \cite{Downie:2014qna} and AMBER at CERN \cite{Adams:2018pwt}. Both experiments were
triggered by the abovementioned proton radius discrepancy. In Ref.~\cite{Kaiser:2022pso}
the radiative corrections to $\mu^\mp p$ scattering specifically
for the kinematics of the AMBER experiment, which operates with a high-energetic muon beam at $100\,$GeV
and measures in near forward directions, thus spanning the momentum transfers  $32\,$MeV$\,< Q < 141\,$MeV,
were worked out. This momentum transfer range  nicely overlaps with the range of the
upcoming MUSE experiment at PSI with $45\,$MeV$\, < Q< 265$~MeV, 
the PRAD-II experiment at Jefferson Lab for $e^-p$ scattering with $14\,$MeV$\, < Q
< 245$~MeV~\cite{PRad:2020oor} as well as the MAGIC $e^-p$ experiment at Mainz, that aims at a
momentum range $10\,$MeV$\,<Q<292$~MeV~\cite{Denig:2016tpq}.
The AMBER experiment intends to measure the proton radius with an accuracy of better than $0.01$~fm,
which requires a detailed study of the radiative corrections to be able to achieve such an accuracy.
For related work on radiative corrections to $\mu p$ scattering,
see Refs.~\cite{Tomalak:2015hva,Tomalak:2018jak,Peset:2021iul}.

\begin{figure}[t!]\centering
\centering\includegraphics[width=0.45\textwidth]{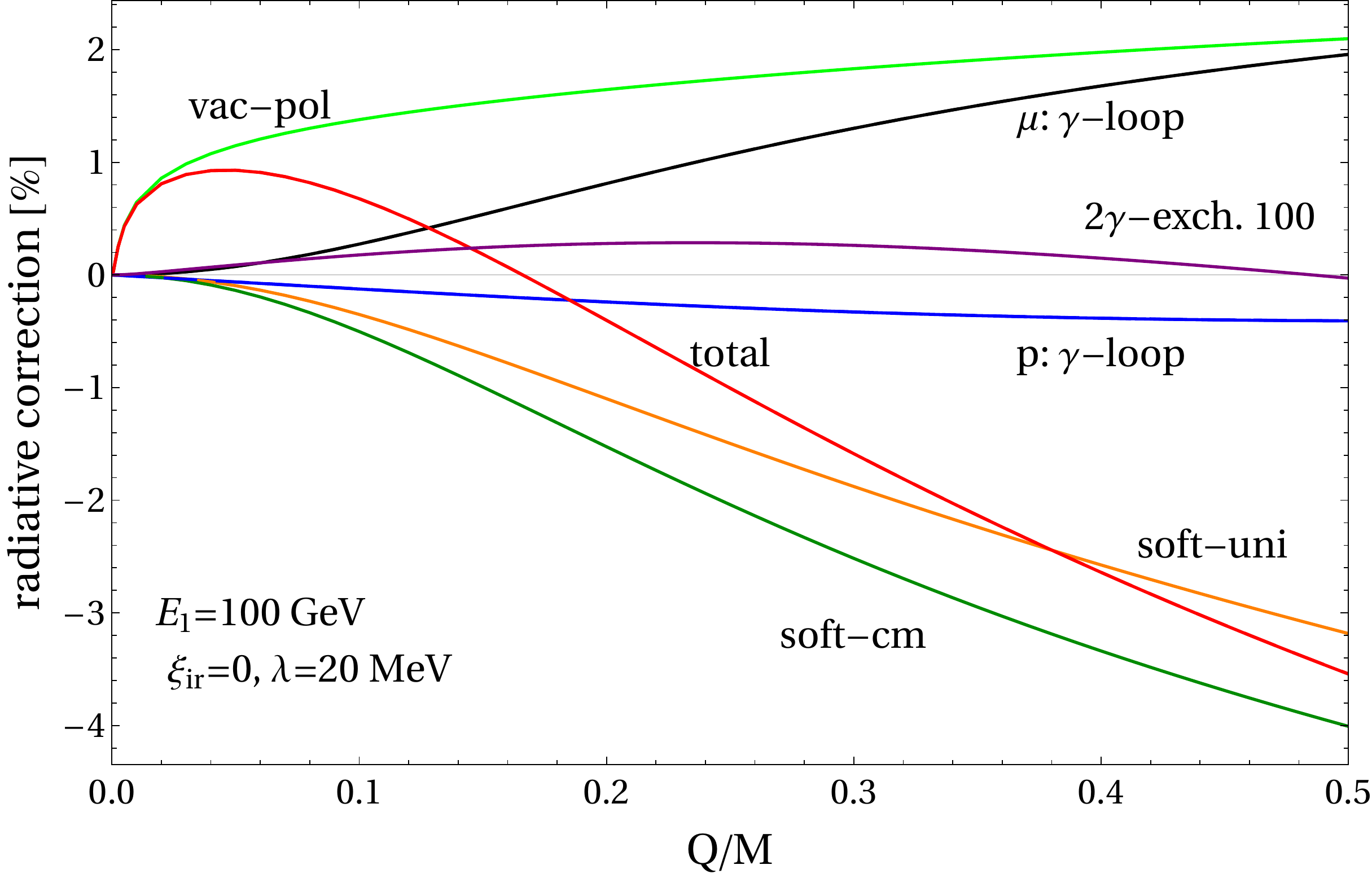}~~~
\includegraphics[width=0.45\textwidth]{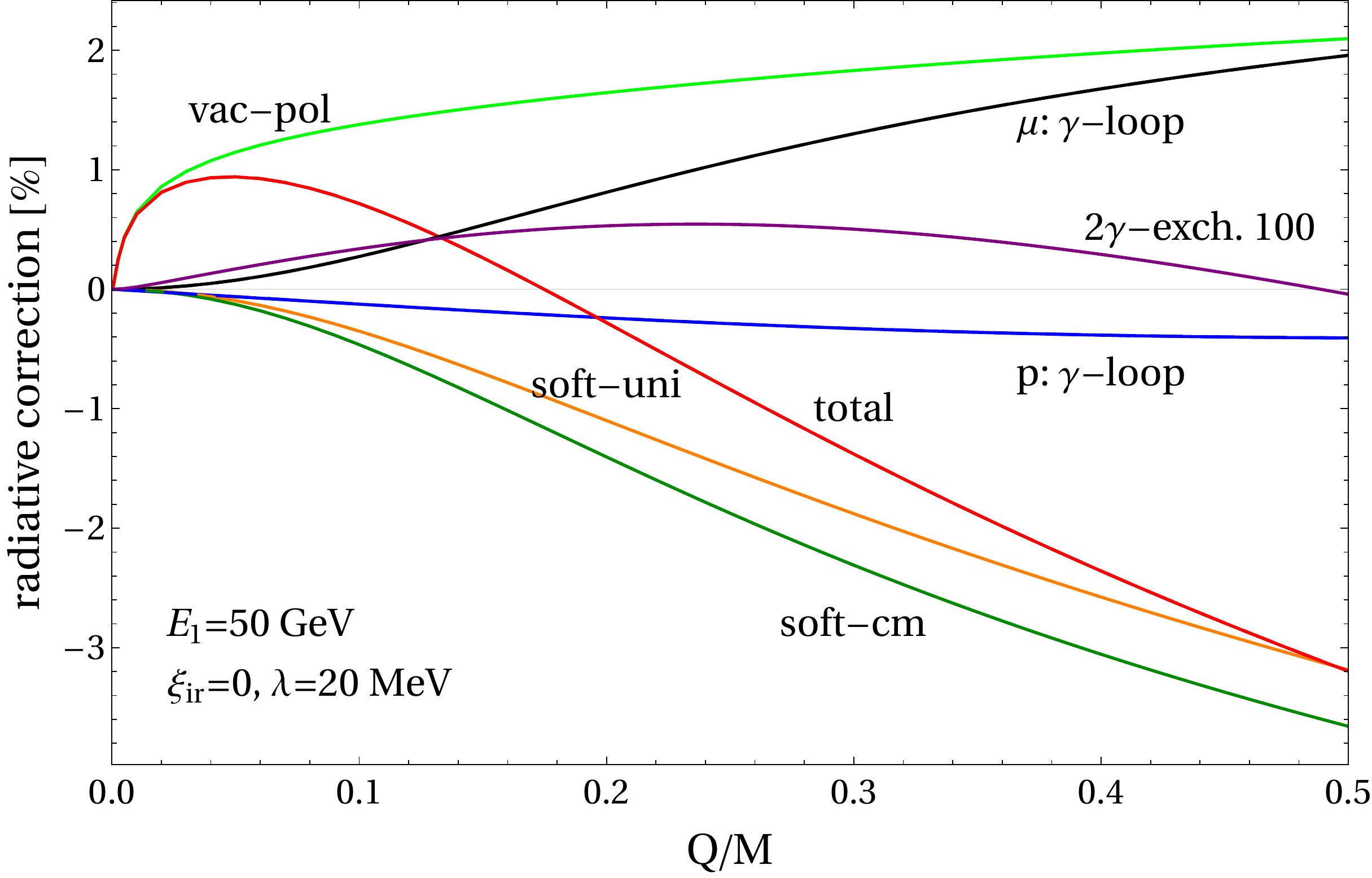}  
\caption{Radiative corrections for the AMBER kinematics with $E_1=100\,$GeV (left panel)
  and $E_1=50\,$GeV (right panel). The individual
  radiative corrections from vacuum polarization, the virtual photon-loops and soft bremsstrahlung are
  shown together with their sum, see the solid line labeled ``total''.}
\label{fig:radcorr100}
\vspace{-3mm}
\end{figure}

Based on earlier works~\cite{Kaiser:2010zz,Kaiser:2016tbf}, Ref.~\cite{Kaiser:2022pso} provided an
update on the radiative corrections for the AMBER experiment based on the high-precision
form factors from~\cite{Lin:2021xrc}. In Fig.~\ref{fig:radcorr100} the radiative corrections from all
sources of order $\alpha/\pi$ for the planned AMBER experiment with a muon beam energy of $E_1 =100\,$GeV
and for $E_1=50\,$GeV are shown,  assuming an infrared cutoff of $\lambda =20\,$MeV. At small
momentum transfers $Q/M \lesssim 0.06$, vacuum polarization is the most dominant effect, because it
is driven by the
electron mass scale. After that, the soft-photon radiation takes over, with a sizable contribution
(of $2\%$) from the photon-loop form factor $2F_1^{\gamma-{\rm fin}}$, involving the muon mass scale,
at the upper end of the momentum transfers considered here. The negative photon-loop form factor
contribution from the proton stays below $0.4\%$ in magnitude, and the two-photon exchange
correction of maximal size $0.3\cdot 10^{-4}$ can essentially be neglected.
It can be seen that for the ratio to the Born cross section, only
the minor terms from the photon-loop form factors of the proton and two-photon
exchange depend on the proton structure predetermined by the strong interactions.
Since a prominent role among the radiative corrections is played by soft photon
radiation, the calculation of the bremsstrahlung process $\mu p \to\mu p \gamma$
should be extended beyond the soft photon approximation and tailored to the
specific experimental conditions. Such work is under way.

\section{Final thoughts}

DRs are arguably the best tool to analyze the electromagnetic form factors of the nucleon
as they allow for a consistent description of all data in the space- and timelike regions
based on fundamental principles. DR analyses
always led to a small proton charge radius and a slightly bigger magnetic one. Also, most
recent experiments tend to the small radius, so the attention has turned from a puzzle
to precision~\cite{Hammer:2019uab}. Also consistently over time, the neutron magnetic
radius has been and is the largest of the electromagnetic nucleon radii. Even more precise
measurements of $ep$ scattering at Jefferson Lab and Mainz are eagerly awaited for as well
as refined lattice QCD calculations. Theory challenges are the consistent extraction of the
neutron form factors from light nuclei along the lines of Ref.~\cite{Filin:2020tcs} and
to obtain a better understanding of the oscillations in the effective form factors
$|G_{\rm eff}^{p,n}|$. Experimental challenges are a measurement of the proton form factor
ratio at $Q^2 \simeq 10\,$GeV$^2$, more resolved form factor measurements in the timelike region
and the investigation of $\mu p$ scattering (MUSE, AMBER) to test lepton flavor universality.

\section*{Acknowledgments}
First, I would like to thank the organizers of INPC 2022 for their superb job done.
Second, I would like to thank my collaborators Yong-Hui Lin, Hans-Werner Hammer,
Norbert Kaiser and Aldo Antognini for sharing their insights into the topics discussed here.
This work  is supported in part by  the DFG (Project number 196253076 - TRR 110)
and the NSFC (Grant No. 11621131001) through the funds provided
to the Sino-German CRC 110 ``Symmetries and the Emergence of
Structure in QCD", by the European Research Council (ERC) under the
European Union's Horizon 2020 research and innovation programme (EXOTIC, grant agreement No. 101018170),
by the Chinese Academy of Sciences (CAS) through a President's
International Fellowship Initiative (PIFI) (Grant No. 2018DM0034) and by the VolkswagenStiftung
(Grant No. 93562).

\end{document}